\documentclass[
reprint, 
showpacs,
nobibnotes,
 amsmath,amssymb,
 aps, 
pra,
]{revtex4-1}

\usepackage{graphicx}
\usepackage{dcolumn}
\usepackage{bm}
\usepackage{feyn}
\usepackage{subfigure}
\usepackage{natbib}
\usepackage{float}
\usepackage{color}
\usepackage{hyperref}
\usepackage{mathrsfs}
\usepackage{rotating}
\usepackage{epsfig}

\newcommand{\be}{\begin{eqnarray}}
\newcommand{\ee}{\end{eqnarray}}
\newcommand{\nn}{\nonumber}
\begin{document}

\title{Structured light beams constituted of incoming and outgoing waves}

\author{Job Mendoza-Hern\'andez}
\email{job.mendoza@alumno.buap.mx}
 \altaffiliation[Present affiliation:
]{Tecnol\'ogico de Monterrey, Escuela de Ingenier\'ia y Ciencias, Ave. Eugenio Garza Sada 2501, Monterrey, N.L.64849, (M\'exico).}
\author{Maximino Luis Arroyo-Carrasco}%
\affiliation{%
Facultad de Ciencias F\'isico -Matem\'aticas. 
Benem\'erita Universidad Aut\'onoma de Puebla, C.P. 72570, Puebla, Pue., M\'exico. 
}%

\author{Marcelo David Iturbe-Castillo}

\author{Sabino Ch\'avez-Cerda}
\email{sabino@inaoep.mx}
\affiliation{
 Instituto Nacional de Astrof\'isica, \'Optica y Electr\'onica,
 Luis Enrique Erro N. 1, 72840 Tonantzintla, Puebla, M\'exico 
}%


\date{\today}
\begin{abstract}

In the present work, we demonstrate that structured light beams are constituted by two traveling waves which transverse components are in opposite directions, that is, incoming and outgoing from the axis of propagation. These waves result from the complex sum of the two fundamental solutions of the transverse component of the spatial wave equation.
After a partial obstruction of the beam, the incoming and outgoing waves can be easily observed during the self-healing process, providing a simple explanation of the phenomenon. Incoming and outgoing waves in Laguerre-Gauss beams are investigated analytically, numerically and experimentally. The proposed way to describe light beams might offer new insights into the phenomenon of diffraction.

\end{abstract}

\maketitle


\section{\label{sec:intruduction}Introduction}

The recent Nobel Prize in optical tweezers reflects the importance of structured light beams \cite{Ashkin1, Ashkin2,Ashkin3,Ashkin4}, beams whose optical properties are specifically tailored according to the application \cite{Allenbk,Torres,Andrews}. 
It has been demonstrated that some structured light beams (SLBs) have orbital angular momentum \cite{Allen}, a characteristic that increases its range of applications to fields like classical and quantum communications \cite{roadmap,Pagett,Willner,Willner1}, imaging \cite{Padgett1}, microscopy \cite{Betzig,Rohrbach}, and micro-manipulation \cite{roadmap}.
The spatial and temporal properties of SLBs are studied with the wave equation and their approximations, with the Helmholtz equation (HE) or the paraxial Helmholtz equation (PHE) \cite{Boyd,siegman,saleh}. 

Since the appearance of lasers, several types of structured light beams have been generated and described, like the Gauss, Hermite-Gauss, Laguerre-Gauss, and more recently Ince-Gauss
 \cite{siegman,ince} and the so-called elegant beams, SLBs with transverse structures described by polynomials with a complex argument \cite{eHermite, eince}. 
 All these beams have distributions that are exact solutions of the PHE \cite{Boyd,GBur} that suffer diffraction during propagation.
 
In the late eighties, the non-diffracting beams appear, e. g., Bessel, Parabolic and Mathieu beams, as new  solutions of the HE \cite{durnin,sabinobessel,mathieu,parabolic}. Furthermore, there are beams which propagate in parabolic trajectories known as Airy beams \cite{airy_berry,airychri}, beams whose complex amplitudes are proportional to the confluent hyper-geometric function \cite{hypermodes, circular}, and beams that do not have an exact solution to the HE or PHE, such as the caustic \cite{jobjmo1} or the Pearcy beams, which describe the diffraction of a caustic cuspid \cite{pearcy}. 

An intriguing property of SLBs is their ability of self-healing \cite{durnin,bouchal, anguiano}, characteristic that is maintained even at the quantum level \cite{McLaren, Eileen}. It occurs when the beam is partially blocked with an opaque object and, after certain propagation distance, it recovers its transverse intensity distribution. The self-healing phenomenon was initially observed in nondiffracting beams\cite{bouchal} and have been described in terms of conical waves \cite{sabinobessel,chavezMc,anguiano,sabinoairy}, by Babinet’s principle \cite{bouchal,agarwal,aiello} and through the relation between two orthogonal field components with an attenuated factor \cite{victorsh}. More recently, this property was investigated in Laguerre-Gauss beams through a comparison with Bessel beams \cite{jobol}.

To describe the internal structure of diffracting and nondiffracting beams only the first solution of the HE and the PHE it is needed, however, a general solution of such equations requires not only one but two solutions.
Keeping this in mind, the aim of this article is to probe that structured light beams are composed of two traveling waves. Two fundamental solutions of the spatial differential equation of second order that have their transverse vectors in opposite directions and its sum generate a structured light beam as a standing wave. In the following, we will refer to such waves as the incoming and outgoing waves. We will demonstrate that Laguerre-Gauss beams contain incoming and outgoing waves, through which it is possible to explain their self-healing ability and show that they are the complex sum of two solutions of the associated Laguerre differential equation of the transverse distribution. The analytic, numerical and experimental features of the incoming and outgoing waves are investigated.

The article is structured as follows: first, we show the representation of the general solution of the wave equation and the spatial wave equation, in terms of two solutions. Then, we derive the second solution of the associated Laguerre differential equation that gives form to the incoming and outgoing waves. Finally, we present the experimental generation of both waves, and when they are partially blocked.

\section{\label{sec:travel}Traveling waves in structured light beams}

The homogeneous wave equation, Eq. (\ref{waveeq}), of an homogeneous medium in regions free of currents and charges, is satisfied by a traveling wave $V(\textbf{r},t)$ \cite{wolf,saleh}.   
\begin{equation}
\nabla^{2}V- \frac{1}{v^{2}}\frac{\partial^{2}V}{\partial t^{2}}=0.
\label{waveeq}
\end{equation}

The simplest solution of this equation is a traveling plane wave $V=V(\textbf{r.s},t)$, with a general solution of the form
\begin{equation}
V=V_{1}(\textbf{r.s}-vt)+V_{2}(\textbf{r.s}+vt),
\label{firtssol}
\end{equation}
where $V_{1}$ and $V_{2}$ are two solutions to the equation. These waves  propagate with velocity $v$ and have a direction opposite  to the plane $\textbf{r.s}=d$, with $d$ a constant.

If the wave can be separated as $V(\textbf{r},t) =R[U(r)F(t)]$, where $R$ denotes the real part and $F(t)=exp(-iwt)$ is an harmonic function, we find that $U$ must satisfy the Helmholtz equation  
\begin{equation}
\nabla^{2}U+k^{2}U=0,
\label{eqhelmholt}
\end{equation}
here $k=w/v$ is the wave vector and $U(\textbf{r})=A(\textbf{r})exp(ig(r))$ is a complex amplitude. Equation (\ref{eqhelmholt}) is the spatial contribution and the general solution depends on the partial differential equation (PDE) that defined the system. 

In order to see the spatial contribution of Eq. \ref{eqhelmholt}, we can use the method of separation of variables to split the PDE into  second-order ordinary differential equations (ODEs). 

Now, we can assign a constant and construct a general solution for each equation. 
In a direct analogy with the wave equation, the spatial wave equation has two solutions with the transverse components of the wave vector in opposite directions. 
Both waves travel in the direction of vector $k$ incoming or outgoing direction from the axis of propagation, hence the name of incoming and outgoing waves.
These waves can be written as 
\begin{equation}
U_{in}=U_{1} - iU_{2},
\label{incoming}
\end{equation}
\begin{equation}
U_{out} =U_{1} + iU_{2},
\label{outgoing}
\end{equation}
where $U_{in}$ and $U_{out}$ are the incoming and outgoing waves, and $U_{1}$ and $U_{2}$ are the two solutions to the second-order ODE. The standing wave is  
\begin{equation}
U =U_{in}+ U_{out}.
\label{eqsuperposition}
\end{equation}
If the Helmholtz equation  is written with circular cylindrical symmetry, and the field is separable in the transverse and longitudinal coordinates as $U(\textbf{r})=U(r)exp(ik_{z}z)$, the incoming and outgoing waves respect to $z$ axis have of the form:
\begin{equation}
U_{in}(r)=U_{1}(r) - iU_{2}(r),
\label{incoming}
\end{equation}
\begin{equation}
U_{out}(r) =U_{1}(r) + iU_{2}(r).
\label{outgoing}
\end{equation}
In this case, $U_{in}(r)$ and $U_{out}(r)$ are Hankel functions \cite{Arfken}, and the standing wave  $U(r)=U_{in}(r)+U_{out}=2U_{1}(r)$ is the first solution of the second-order ODE. 
The simplest case of circular cylindrical symmetry is when $U(r)=2J_{0}(k_{t}r)$, where $J_{0}$ is the Bessel beam and  $k^{2}=k_{z}^{2}+k_{t}^{2}$  with $k_{t}$ the transverse component and $k_{z}$  the longitudinal. Since the intensity is independent of $z$, $I(r,z)=I(r,0)$, these beams are known as nondiffracting beams \cite{durnin}, showing the self-healing property \cite{sabinobessel,anguiano}.

On the other hand, diffracting beams $I(r,z)$ are analyzed with the paraxial Helmholtz equation \cite{ siegman,saleh}, a second-order PDE whose solutions can be found with the method of separation of variables. In circular cylindrical symmetry, these solutions are 
\begin{equation}
U_{in}(r,z)=U_{1}(r,z) - iU_{2}(r,z),
\label{incomingpa}
\end{equation}
\begin{equation}
U_{out}(r,z) =U_{1}(r,z) + iU_{2}(r,z),
\label{outgoingpar}
\end{equation}
here $U_{in}(r,z)$ and $U_{out}(r,z)$ are the incoming and outgoing waves for the paraxial approximation, and $U_{1}(r,z)$ and $U_{2}(r,z)$ are the two linearly independent solutions. Thus, the standing wave for diffracting beams is 
\begin{equation}
U(r,z)=U_{in}(r,z)+ U_{out}(r,z).
\label{eqsuperpositionparaxial}
\end{equation}
Equation (\ref{eqsuperpositionparaxial}) gives a shape structured by two waves.
Therefore, we can easily observe the behavior of the two beams when they are propagated and also partially obstructed.

\section{\label{sec:Eq} The construction of incoming and  outgoing waves for Laguerre-Gauss beams}

The traveling waves  that describe the propagation of Laguerre-Gauss beams (LGBs) are found with the solutions of the associated Laguerre differential equation (ALDE). 

It is well-known that LGBs are solutions of the paraxial wave equation in cylindrical coordinates $(r,\varphi,z)$ \cite{siegman,GBur}. The ALDE equation that governs the radial features of the beam, is obtained through the method of separation of variables and its first solution is given by the \textit{associated Laguerre polynomial}  $L_{n}^{m}(x)$ \cite{Arfken,Agarwal:2009,Trench:2000,Nagle:2008} (for simplicity we will only use the $x$ coordinate). The parameters $n$ and $m$ are the radial and azimuthal orders of the polynomial have integer values.
It is important to highlight that the second solution where $n$ and $m$ integer values is not present in the literatu. In this document, we indicate a way to build it using the first solution.

 In order to find the second solution with $n$ and $m$ integer values, it is useful to use the confluent hypergeometric function (CHF) of first kind $\Phi(a,m;x)$ with $a$ non-integer number \cite{Arfken,Bell}. Then the associated Laguerre polynomial $L_{n}^{m}(x)$ \cite{Arfken} can be written as
\begin{equation}
L_{n}^{m}(x)= \frac{(m+n)!}{m!n!} \Phi(-n;m+1,x),
\label{polylacon}
\end{equation}
which gets reduced to the associated Laguerre polynomial  of degree $n$ when the index of the sum is $k=n$, and also when multiplied by  $(m+n)!/m!n!$.

A second solution to ALDE is introduced with through function $U(a,m,x)$, the confluent hypergeometric function of the second kind (Tricomi function) \cite{Arfken}.
The problem with $\Phi(-a,m+1;x)$ and $U(-a,m+1;x)$ is that they are dependent functions when m=0,1,2,.. and a=n=0,1,2,....
\be
{U(-n,m+1,x)}& = &(-1)^{n}\frac{(m+n)!}{m!}\Phi(-n,m+1,x).
\label{depend}
\ee
However,  $\Phi$ and $U$ can be used to define a second solution to ALDE, as we will see in the next section.

\subsection{\label{sec:Eq} X function as second solution and Laguerre-Hankel functions. Analytic form of the incoming and outgoing waves}

We define the second solution of the ALDE for $m$ and $n$ integers following the methodology given by Hankel and Weber to find the second solution of the  Bessel differential equation \cite{Watson}. After reviewing the math tables \cite{gradshteyn,abramowitz}, we have decided to call this new function $X$.

The function $X$ is defined as \cite{tesisjob}:
\be
{X_{a}^{m}(x)}& = & \frac{D U\cos(a\pi)-\Phi}{\sin(a\pi)},
\label{lnnoentero}
\ee
where $a$ is not an integer, $m=0,1,2$..., and  $D=\frac{m!}{(m+a)!}$. 
For simplicity, we write $U$ and $\Phi$ instead of $U(-a,m+1;x)$ and $\Phi(-a,m+1;x)$, respectively. 
The function $X$ for the integer indexes $m$ and $n$ is constructed as a Neumann function \cite{Arfken}. When the value of $a \rightarrow n$, the right-hand side of Eq. (\ref{lnnoentero}) becomes undefined, and it approaches a limit. Using  L'Hopital's rule, we found  that \cite{tesisjob}:
\be
{X_{n}^{m}(x)}&  =& \lim_{a \rightarrow n}X_{a}^{m}(x) \nn\\&&=\frac{1}{\pi}\left[(-1)^{n}\frac{\partial(D U)}{\partial a}-\frac{\partial(\Phi)}{\partial a}\right]_{a=n},
\label{lnentero}
\ee
where $X_{n}^{m}(x)$ is the second solution of the ALDE for $n=0, 1, 2...$ and $m=0, 1,2,...$. The factor $D=\frac{m!}{(m+a)!}$ is necessary to match amplitudes and  compare $X_{n}^{m}(x)$ with the function $\Phi(-n,m+1,x)$. Then, the  ALDE's  most general solution is given by:
\be
y(n,m;x)=A\Phi(-n,m+1;x)+BX_{n}^{m}(x).
\label{sologeneral}
\ee
Here $A$ and $B$ are constants, and the indexes $m$ and $n$ are integers. 

If we multiply $\Phi$ and $X$ from Eq. (\ref{sologeneral}) by
\be
C(m,n)=\frac{(m+n)!}{m!n!},
\ee
then  $X_{n}^{m}=CX_{n}^{m}$ and $L_{n}^{m}=C\Phi(-n,m+1;x)$.

The solutions $X_{n}^{m}$ and $L_{n}^{m}$ of the ALDE allow us to define the following Hankel type functions  
\begin{equation}
	LH_{n}^{m}(1;x)=L_{n}^{m}(x) + i X_{n}^{m}(x),
	\label{laguerrhankel1}
\end{equation}
\begin{equation}
	LH_{n}^{m}(2;x)=L_{n}^{m}(x) - i X_{n}^{m}(x).
	\label{laguerrhankel2}
\end{equation}
 Therefore, Eq. (\ref{laguerrhankel1}) and Eq. (\ref{laguerrhankel2}) are \textbf{Laguerre-Hankel} functions of  first and second kind. When we have both functions simultaneously, the result is a Laguerre polynomial:
\begin{equation}
	[LH_{n}^{m}(1;x)+LH_{n}^{m}(2;x)]=2L_{n}^{m}(x). 
	\label{laguerrbyadd}
\end{equation}
The Laguerre-Hankel functions describe the two fundamental functions, in the transverse direction, for the generation of incoming and outgoing waves. 

\section{\label{sec:Eq1} Behavior of incoming and outgoing waves in Laguerre-Gauss beams}

We consider the definition of Laguerre-Gauss beams as in \cite{GBur}. The incoming and outgoing waves that constitute a Laguerre-Gauss beam are defined by the Laguerre-Hankel functions $LH_n^{|m|}$ described as 
\begingroup
\begin{multline}
  U_{in,out}{=}A_0\frac{w_0}{w(z)}\left(\frac{\sqrt{2}r}{w(z)}\right)^{|m|}
  LH_{n}^{|m|}\left(p; \frac{2r^{2}}{w^{2}(z)}\right)\times\\
  \exp\left[ {\frac{-r^{2}}{w^{2}%
        (z)}-\frac{ikr^{2}}{2R(z)}+i(2n+|m|+1)\Phi(z)- im\varphi}
  \right],
  \label{laguerresol}
\end{multline}
\endgroup
where $p=1,2$, and $n$ and $m$ are the radial and azimuthal indices. Also, it is possible defined the X-Gauss beam (XGB) substituting $X_n^{|m|}$ by $LH_n^{|m|}$ in the Eq. \ref{laguerresol}, in order to comparison between the two solutions of ALDE and the incoming and outgoing waves.   

The parameters in Eq. (\ref{laguerresol}) are the beam waist $w_{0}$,  beam width $w^{2}\left( z\right)=w_{0}^{2}\left[ 1+\left( z/L_{D}\right)
  ^{2}\right] $,  transverse phase front $R\left( z\right) =z\left[
  1+\left(L_{D}/z\right) ^{2}\right] $ and the Gouy phase shift
$\Phi(z)=\tan^{-1}(z/L_D)$. In all these expressions $L_{D}=kw_0^2/2$ is the Rayleigh distance or diffraction distance for the wave number $k=2\pi/\lambda$.

In order to show the features of the incoming and outgoing waves, we have normalized the transverse coordinate to $w_{0}$ and the longitudinal coordinate $z$ to $L_{D}$ in Eq. (\ref{laguerresol}). 
It is important to note that the explicit solution of $X_{n} ^{m}$, as well as its complete features, will be reported elsewhere.

The purpose of this work is not only to observe the behavior of the LGB in terms of two constituent waves, the incoming and the outgoing, but also to find a method to obtain them.
Therefore, we approach the function $X_{n} ^{m}$ by taking the nearest integer as $a\rightarrow n$ in the first term of the sum in Eq. (\ref{lnentero}), taking $k=n$ elements of the series. 
In the second term of the sum, we take at least $k=3n+1$ elements of the series, this in order to have the same number of rings in the XGB and the LGB. In this way, we avoid the divergence on both sides of the function. 

\begin{figure}[H] \centering
\includegraphics[width=8cm]{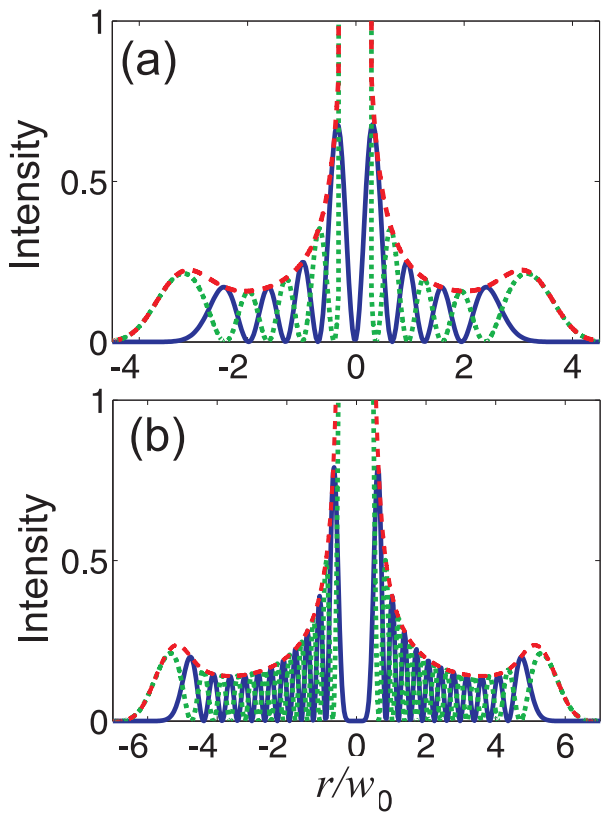}
\hspace{0.1cm}
\caption{\footnotesize{Intensity profile of the incoming-wave or outgoing-wave (red), LGB (blue), and  XGB (green). (a)  the case $m=1$ and $n=3$, (b) $m=5$ and $n=10$.}} \label{centralprom0n10}
\end{figure}

Figure \ref{centralprom0n10} shows the central intensity profile of the incoming wave, including the LGB and the XGB. 
The incoming and outgoing waves diverge at the origin because of their logarithmic solutions, since the function $X$ appears in the second solution \cite{Arfken}. 
We selected the intensity at the center of the beams to be at least three times larger than that of the lobes, in order to distinguish their contributions. 
We can easily observe the relation between teh two beams, 
LGB has a maximum where XGB has a minimum. 
Figure \ref{centralprom0n10} (a) shows the case when $m=1$ and $n=3$, while in Fig. \ref{centralprom0n10} (b) $m=5$ and $n=10$.  

\begin{figure}[H] \centering
\includegraphics[width=8.2cm]{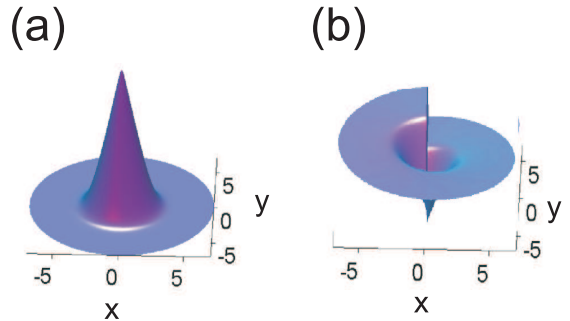}
\hspace{0.1cm}
\caption{\footnotesize{Conical wavefront for incoming and outgoing waves: (a) without orbital angular momentum $m=0$, and (b) with orbital angular momentum $m\neq0$. }} 
\label{wavefront}
\end{figure}

The incoming and outgoing waves have a conical wavefront Fig. \ref{wavefront}, the first traveling incoming with respect to the axis, and the latter, in the opposite direction. 
The function $\alpha_{in,out}$ is plotted for each $U_{in,out}$. 
Figure \ref{wavefront} shows the wavefront for $z=0$, and $t=0$. 
Figure  \ref{wavefront}(a) the case $m=0$ and $n=3$, and Fig. \ref{wavefront}(b) shows the case $m=1$ and $n=3$, mixing the conical and helical wavefronts.

\begin{figure}[H] \centering
\includegraphics[width=6.5cm]{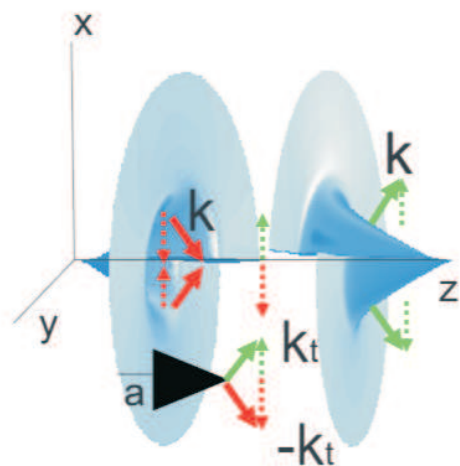}
\hspace{0.1cm}
\caption{\footnotesize{Wavefronts of the incoming and outgoing waves: direction of the transverse wave vector after an obstruction of radius $a$.}} 
\label{selfhealingwf}
\end{figure}

The wavefronts of the incoming and outgoing waves help us to understand the self-healing behavior. 
If the beam is partially blocked by an obstruction of radius $a$, the beam will recover its transverse profile due to the sum of these two waves. 
The opposite directions of the perpendicular wave vectors hold the stationary distribution after the obstruction (see Fig. \ref{selfhealingwf}). 
The LGB has a conical wavefront, although there is a curve in the conical base, as shown in Fig. \ref{selfhealingwf}. 
For the other transverse distributions, the symmetry of the beam may give other surfaces; however, there will be two perpendicular wave vectors with opposite directions to hold the shape of the beam. 

\begin{figure}[H] \centering
\includegraphics[width=8cm]{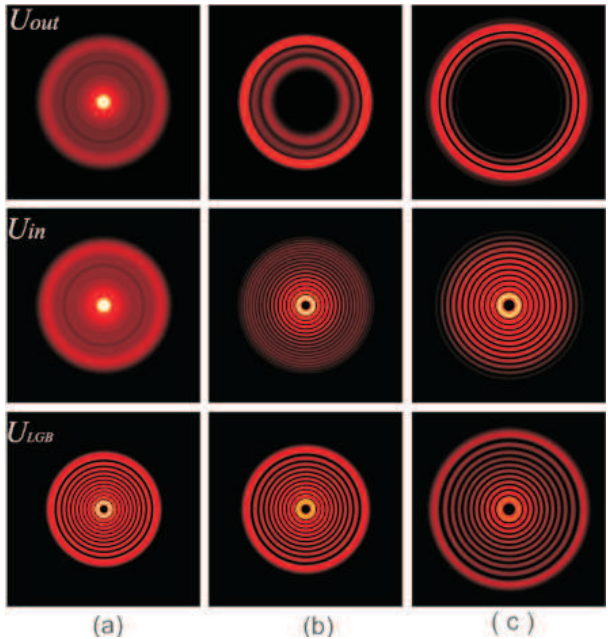}
\hspace{0.1cm}
\caption{\footnotesize{Stages of propagation of incoming and outgoing waves and LGB: (a) with $m=5$ and $n=10$ at the position $z=0$, (b) $z=0.5L_{D}$, and  (c) $z=L_{D}$. The incoming wave is shown in the first row, the outgoing-wave  in the second row, and the LGB in the last row.  }} \label{behavioratz}
\end{figure}

The behavior under propagation of the incoming and outgoing waves is shown in Fig. \ref{behavioratz}. 
The outgoing wave generates the external part of the LGB as shown in the first row of the figure. 
The incoming wave generates the internal part of the LGB; this feature acts as an axicon for the generated LGB as in the case of Bessel beams\cite{anguiano}. Thus, it is possible to consider the generation of optical elements using the incoming wave.

\section{\label{sec:Eq1}Experimental generation of incoming and outgoing waves}

\begin{figure}[htp] \centering
\includegraphics[width=8.5cm]{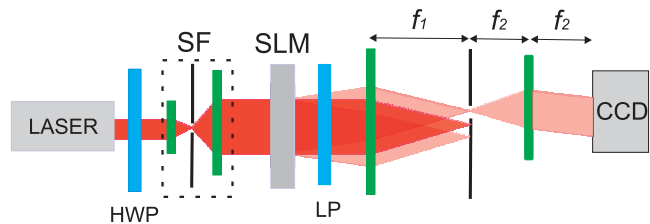}
\hspace{0.1cm}
\caption{\footnotesize{ Experimental scheme for the generation of incoming and outgoing waves ($U$) with a computer-generated hologram.  $HWP$: half wave plate, $SF$: filter system, $P$: polarizer, $SLM$: spatial light modulator, $f_{1}=38cm$ and $f_{2}=10cm$ lenses, and  $CCD$: camera.  }} \label{setup}
\end{figure}

We have generated experimentally the incoming and outgoing waves and observed these waves when a Laguerre-Gauss beam is partially blocked. We used the technique developed by Arriz\'on  \textit{et al.} \cite{victor}, as shown in Figure \ref{setup}. 
A computer-generated hologram (CH) was implemented into an amplitude liquid crystal spatial light modulator SLM (HOLOEYE-LC2002). The light source is a He-Ne laser (632.8nm) with polarization parallel to the director axis of the SLM. A polarizer is placed after the SLM to modulate the CH and eliminate an extra phase. The beam is obtained with a combination of a 4f-system and an aperture located at the Fourier plane to decode the wave $U$. The CCD camera keeps the beam after passing through the second lens $f_{2}$. 

 
\begin{figure}[H] \centering
\includegraphics[width=8.5cm]{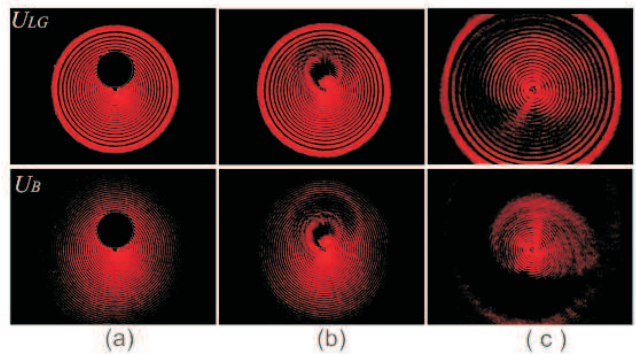}
\hspace{0.1cm}
\caption{\footnotesize{The self-healing comparison between a Laguerre-Gauss beam  with azimuthal index $m=5$ and radial index $n=19$ and a Bessel beam, both partially obstructed at the position: (a) $z=0cm$, (b)  $z=8cm$ , and (c) $z=34cm$.  We can see the behavior of two shadows due to the constitutive waves of the beam in (b). }} \label{lgvsBessel}
\end{figure}

We can observe qualitatively the behavior of the incoming and outgoing waves in the partially blocked Laguerre-Gauss beams while they are self-healing, as shown in Fig. \ref{lgvsBessel}. We also observe the behavior on Bessel beams with the same transverse section and spatial frequency for comparison, as in the work done by Mendoza-Hern\'andez \textit{et al.}\cite{jobol}.  
Figure \ref{lgvsBessel}(a) shows the LGB in the first row and Bessel beam (BB) in the second row, partially blocked. Figure \ref{lgvsBessel}(b) shows two shadows due to the incoming and outgoing waves: a large low contrast shadow due to the outgoing wave and a small high contrast shadow due to the incoming-wave. 
The diffracted shadow due to the incoming wave affects the innermost ring  and moves around the center of the pattern, until it becomes an outgoing wave that increases its size and moves away from the center (Fig. \ref{lgvsBessel}(c)). Using a lens  $f_{2}=10cm$ into Fig. (\ref{setup}), the LGB has a waist $w_{0}=0.2644mm$ and a diffraction distance of $L_{D}= 34.69cm$. Figure \ref{lgvsBessel}(c) shows the self-healing of LGB and BB.

\begin{center}
\begin{figure}[H] \centering
\includegraphics[width=8.5cm]{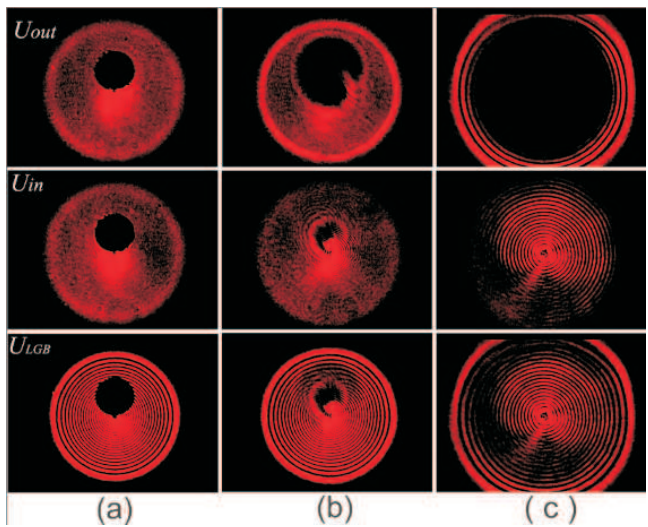}
\hspace{0.1cm}
\caption{\footnotesize{ Incoming and outgoing waves, and LGB partially obstructed with  $m=5$ and $n=19$ at the position: (a) $z=0cm$,  (b) $z=8cm$, and (c) $z=34cm$.}} \label{exoticexpe}
\end{figure}
\end{center}

Figure \ref{exoticexpe} shows the behavior of the partially obstructed incoming and outgoing waves, and the partially blocked LGB. It portrays the dynamics of self-healing contained in LGB. 
The first row presents the large shadow due to the outgoing wave that diffracts away from the center. 
In the second row, a high contrast shadow is seen due to the incoming wave that affects the innermost ring, which moves around the center. 
Therefore, the behavior in the partially blocked LGB is caused by two waves that hold the stationary transverse distributions.

\section{Discussion and Conclusion}
It is well known that a solution of the wave equation is a traveling wave. In the Helmholtz equation, a second-order differential equation that describes the spatial features of a beam, has two solutions; however, only the first one is necessary to describe the internal structure of diffracting and nondiffracting beams. In this work, we have demonstrated that structured light beams can also be described through both solutions, two traveling waves with the particularity of having their transverse vector in opposite directions, waves that we have called the incoming wave and the outgoing wave. Since the standing wave corresponds to the sum of the incoming and the outgoing waves, it is possible to obtain information about each wave when the beam is partially blocked during propagation.

The characteristics of the incoming and outgoing waves in the Helmholtz equation and its paraxial approximation were studied. In particular, we have demonstrated the existence of incoming and outgoing waves in Laguerre-Gauss beams, treated as diffracting beams. First, it was necessary to find the second solution of the associated Laguerre differential equation for integer radial and azimuthal indexes or function $X$, as we named it. Then, function $X$ was used to construct the Laguerre-Hankel functions of first and second order that would define the incoming and outgoing waves. The generation and propagation of such waves were presented numerically and experimentally, showing that the diffracting spreading due to propagation does not affect observed their fundamental components. 
Finally in conclusion, a structured beam can be seen as a standing wave that results from the sum of the incoming and outgoing waves, and thus, providing an explanation for its self-healing ability.

\begin{acknowledgments}
Job Mendoza-Hern\'andez acknowledges support from the National Council of Science and Technology of Mexico, CONACYT. JMH thank Victor Arrizon for the summer classes about SLMs, and Dorilian López-Mago, Ricardo Tellez-Limón,  Adriana Inclan-Ladino, and Maribel Hidalgo-Aguirre for the fruitful discussions about the present work.

\end{acknowledgments}


\begin{thebibliography}{99}
\bibitem{Ashkin1} A. Ashkin, ``Acceleration and Trapping of Particles by Radiation Pressure,'' Phys. Rev. Lett. \textbf{24}, 156, (1970).

\bibitem{Ashkin2} A. Ashkin, and J.M. Dziedzic, ``Optical trapping and manipulation of viruses and bacteria,'' Science \textbf{235}, 1517-1520 (1987).

\bibitem{Ashkin3} A. Ashkin, J.M. Dziedzic, and T. Yamane,  ``Optical trapping and manipulation of single cells using infrared laser beams,'' Nature \textbf{330}, 769-771 (1987).
\bibitem{Ashkin4} A. Ashkin, J.M. Dziedzic, J. E. Bjorkholm, and S. Chu, ``Observation of a single-beam gradient force optical trap for dielectric particles,'' Opt. Lett. \textbf{11}, 288-290 (1986).


\bibitem{Allenbk} L. Allen, S.M. Barnett, and M.J. Padgett (Eds.), \emph{Optical Angular Momentum} (IOP Publishing, 2003).

\bibitem{Torres} J.P. Torres and L. Torner (Eds.), \emph{Twisted Photons: Applications of Light with Orbital Angular Momentum}, (Wiley-VCH, 2011).

\bibitem{Andrews} D.L. Andrews and M. Babiker (Eds), \emph{The Angular Momentum of Light}, (Cambridge University Press, 2012).

\bibitem{Allen} L. Allen, M. W. Beijersbergen, R. J. C. Spreeuw, and J. P. Woerdman, `` Orbital angular momentum of light and the trasformation of Laguerre-Gaussian laser modes,''
Phys. Rev. A \textbf{45}, 8185, (1992)

\bibitem{roadmap} H. Rubinsztein-Dunlop, A. Forbes, M. V. Berry,
M. R. Dennis, D. L. Andrews, M. Mansuripur, C. Denz,
C. Alpmann, P.Banzer, T. Bauer, E. Karimi,
L. Marrucci, M. Padgett, M. Ritsch-Marte,
N. M. Litchinitser, N. P. Bigelow, C. Rosales-Guzm\'{a}n,
A. Belmonte, J. P. Torres, T. W. Neely, M. Baker,
R. Gordon, A. B. Stilgoe, J. Romero,
A. G. White, R. Fickler, A. E. Willner, G. Xie,
B. McMorran, A. M. Weiner, ``Roadmap on structured
light,'' J. Optics \textbf{19}, 013001 (2017).


\bibitem{Pagett} M. J. Padgett,  ``Orbital angular momentum 25 years on,''  Opt. Express \textbf{25}, 11265 (2017).

\bibitem{Willner} J. Wang, J.Y. Yang, I. M. Fazal, N. Ahmed, Y. Yan, H. Huang, Y. Ren, Y. Yue, S. Dolinar, M. Tur, and A. E. Willner, ``Terabit free-space data
transmission employing orbital angular momentum multiplexing,'' Nature Photonics, \textbf{6}, (2012).

\bibitem{Willner1} K. Pang, C. Liu, G. Xie, Y. Ren, Z. Zhao, R. Zhang, Y. Cao, J. Zhao, H. Song, H. Song, L. Li, A. N. Willner, M. Tur, R. W. Boyd, A. E. Willner, ``Demonstration of a 10 Mbits quantum communication link by encoding data on two LaguerreGaussian modes with diferent radial indices,'' Opt. Lett. \textbf{43}, (2018).

\bibitem{Padgett1} M. J. Padgett, R. W. Boyd,  ``An introduction to ghost imaging: quantum and classical,'' Phil. Trans. R. Soc. A, \textbf{375}, (2017)

\bibitem{Betzig} T. A. Planchon, L. Gao, D. E. Milkie, M. W. Davidson, J. A. Galbraith, C. G. Galbraith, E. Betzig, ``Rapid three-dimensional isotropic imaging of living cells using Bessel beam plane illumination,'' Nature Meth. \textbf{8}, 417 (2011).

\bibitem{Rohrbach} F. O. Fahrbach, P. Simon, A. Rohrbach, ``Microscopy with self-reconstructing beams,'' Nat. Phot. \textbf{4}, 780-785 (2010).

\bibitem{Boyd}G. D. Boyd, H. Kogelnik, ``Generalized Confocal Resonator Theory,'' Bell
Syst. Tech. J. 41, 1347-1369 (1962).

\bibitem{siegman}
A.~E. Siegman, \emph{{Lasers}} (University Science Books, 1986).

\bibitem{saleh} B. E. A. Saleh, M. C. Teich, \emph{Fundamentals of Photonics} (John Wiley and
Sons, 1991).


\bibitem{ince}
M. A. Bandres, and J. C. Guti\'errez-Vega, ``Ince-Gaussian beams,'' Opt. Lett. \textbf{29}, 144 (2004).

\bibitem{eHermite}
 A. E. Siegman, ``Hermite- Gaussian functions of complex argument as optical-beam eigenfunctions,'' J. Opt. Soc. Am. \textbf{63}, 1093 (1973).

 \bibitem{eince}
M. A. Bandres, ``Elegante Ince-Gaussian beams,'' Opt. Lett. \textbf{29}, 1724 (2004).

\bibitem{GBur} G. J. Gbur, \emph{Mathematical Methods for Optical Physics and Engineering}  (Cambridge University Press, 2011).


\bibitem{durnin}
J. Durnin, J. J. Miceli, Jr., and J. H. Eberly,  ``Diffraction-Free Beams,'' Phys. Rev. Lett. \textbf{58}, 1499 (1987).

\bibitem{sabinobessel}
S. Ch\'avez-Cerda, ``A new approach to Bessel beams,'' J. Mod. Opt. \textbf{46}, 923 (1999).
\bibitem{mathieu}
 J. C. Guti\'errez-Vega, M. D. Iturbe-Castillo, and  S. Ch\'avez-Cerda, ``Alternative formulation for invariant optical fields: Mathieu beams,'' Opt. Lett. \textbf{25}, 1493 (2000).

 \bibitem{parabolic}
 M. A. Bandres, J. C. Guti\'errez-Vega, and S. Ch\'avez-Cerda, ``Parabolic nondiffracting optical wave fields,'' Opt. Lett. \textbf{29}, 44 (2004).
 
\bibitem{airy_berry}M. V. Berry, N. L. Balazs, ``Nonspreding wave packets,'' Am. J. Phys. \textbf{43},
264 (1979).

\bibitem{airychri} G. A. Siviloglou, J. Broky, A. Dogariu, D. N. Christodoulides, ``Observation
of accelerating Airy beams,'' Phys. Rev. Lett. \textbf{99}, 213901 (2007). 

\bibitem{hypermodes}
V. V. Kotlyar, R. V. Skidanov, S. N. Khonina, and V. A. Soifer, ``Hypergeometric modes,'' Opt. Lett. \textbf{32}, 724, (2007).

\bibitem{circular}
M. A. Bandres, and J. C. Guti\'errez-Vega, ``Circular beams,'' Opt. Lett. \textbf{33}, 177 (2008).

 \bibitem{pearcy} J. D. Ring, J. Lindberg, A. Mourka, M. Mazilu, K. Dholakia, and M. R. Dennis, ``Auto-focusing and self-healing of Pearcey beams,'' Opt. Express,  \textbf{20},  18955 (2012).

\bibitem{jobjmo1} J. Mendoza-Hern\'andez, M. L. Arroyo Carrasco, M. M. M\'endez Otero, S. Ch\'avez-Cerda, M. D. Iturbe Castillo, ``New asymmetric propagation inavariant
beams obtained by amplitude and phase modulation in frequency space,'' J.
Mod. Opt. \textbf{61}, S46-S56 (2014).



\bibitem{anguiano} M. Anguiano-Morales, M. M. M\'endez-Otero, M. D. Iturbe-Castillo, and S. Ch\'avez-Cerda,  ``Conical dynamics of Bessel beams,''  Opt. Eng. \textbf{46}, 078001 (2007).

\bibitem{bouchal} Z. Bouchal, J. Wagner, and M.Chlup,  ``Self-reconstruction of a distorted nondiffracting beam,''  Opt. Commun. \textbf{151}, 207 (1998).

\bibitem{McLaren}M. McLaren, T. Mhlanga, M. J. Padgett, F. S. Roux, and A.
Forbes, ``Self-healing of quantum entanglement
after an obstruction,'' Nat. Commun. \textbf{424}, (2014).

\bibitem{Eileen} E. Otte, I. Nape, C. Rosales-Guzm\'an, A. Vall\'es, C. Denz, and A. Forbes, ``Recovery of nonseparability in self-healing vector Bessel ,'' Phys. Review A, \textbf{98}, 053818 (2018).

\bibitem{chavezMc}S. Chávez-Cerda, G. S. McDonald, G. H. C. New, ``Nondiffracting beams: traveling, standing, rotating and spiral waves,'' \textit{Opt. Commun.} \textbf{123}, 225 (1996).


\bibitem{sabinoairy} J. Rogel-Salazar, H. A. Jim\'enez-Romero, and S. Ch\'avez-Cerda, ``Full characterization of Airy beams under physical principles,'' Phys. Rev. A \textbf{89}, 023807 (2014).


\bibitem{agarwal} J. Aiello, G. S. Agarwal, ``Wave-optics description of self-healing mechanics in Bessel beams,'' \textit{Opt. Lett.}, \textbf{39}, 6819 (2014).

\bibitem{aiello} A. Aiello, G. S. Agarwal, M. Pa\'ur,
B. Stoklasa, Z. Hradil, Jaroslavreh\'acek, P. de la Hoz, G. Leuchs, 
L. L. S\'anchez-Soto, ``Unraveling beam self-healing,'' \textit{Opt. Express},  \textbf{25}, (2017). 

\bibitem{victorsh}
V. Arrizon, G. Mellado-Villaseñor,  D. Aguirre-Olivas,  H. M. Moya-Cessa, ``Mathematical and diffractive modeling of self-healing,'' \textit{Opt. Express},  \textbf{26}, (2018). 

\bibitem{jobol} J. Mendoza-Hern\'andez, M. L. Arroyo Carrasco, M. D. Iturbe Castillo, and S. Ch\'avez-Cerda, ``Laguerre-Gauss beams vs Bessel beams showdown: peer comparison,''  Opt. Lett. \textbf{40}, 3739-3742 (2015). 

\bibitem{wolf} M. Born, E. Wolf,  \emph{Principles of optics. Electromagnetic theory of propagation, interference and diffraction of light.}  (Seventh (expanded) edition, Cambridge University Press, 1999).


\bibitem{Lebedev} N. N. Lebedev, \emph{Special Functions and their Applications}  (Dover Publications Inc., 1972).

\bibitem{Arfken}
G. B. Arfken, and H. J. Weber, \emph{Mathematical Methods for Physicists}  (Elsevier Academic Press, Fifth edition, 2001).


\bibitem{Agarwal:2009}
R. P.~ Agarwal, D. O. Regan, \emph{Ordinary and Partial Differential Equations. With Special Functions, Fourier Series, and Boundary Value Problems} ( Springer, 2009).

\bibitem{Trench:2000} W. F. Trench, \emph{Elementary differential Equations} (Brooks/Cole Thomson Leaning, 2000).

\bibitem{Nagle:2008}
R. K. Nagle, E. B. Saff, and A. D. Snider, \emph{Fundamentals of differential equations} (Pearson Addison Wesley, $7$th edition, 2008).

\bibitem{Bell} W. W. Bell, \emph{Special functions for scientists and engineers}  (Dover Publications Inc.,  2004).

\bibitem{gradshteyn} I. S. Gradshteyn, and I. M. Ryzhik. \emph{Table of Integrals, Series, and Products} (Academic Press, 2007).

\bibitem{abramowitz} M. Abramowitz and I. A. Stegun \emph{Handbook of Mathematical Functions}, (Dover, New York, 1964).

\bibitem{Watson} G. N. Watson, \emph{A Treatise on the Theory of Bessel Function}, (Cambridge at University Press, Cambridge, 1966).

\bibitem{tesisjob} Job Mendoza-Hern\'andez, PhD thesis, Estudio de la Auto - Reconstruccion de Algunos Campos Estructurados de Luz, ``Study of the Self-Healing of some Structured Light Fields''. Benem\'erita Universidad Auton\'oma de Puebla, Facultad de Ciencias F\'isico Matem\'aticas, Puebla, M\'exico  (2016).

\bibitem{victor} V. Arriz\'on, G. Mendez, D. S\'anchez-de-la-llave ``Accurate encoding of arbitrary complex fields with amplitude-only liquid crystal spatial light modulators,'' \textit{Opt. Express} \textbf{13}, 7913-7927 (2005)


\end{thebibliography}
\end{document}